\documentclass[aps,prc,twocolumn,superscriptaddress,preprintnumbers,amsmath,amssymb,floatfix,showpacs]{revtex4}

\usepackage{graphicx}
\usepackage{dcolumn}
\usepackage{bm}
\usepackage{longtable}
\usepackage{color}
\usepackage{float}

\begin{document}

\preprint{MAX-lab $^{3}$He Summary Article for arXiv v3}

\title{A measurement of the differential cross section for the two-body photodisintegration of $\bm{{^3}}$He at $\bm{{\theta^{\rm LAB}=90^\circ}}$ using tagged photons in the energy range 14--31 MeV}

\author{\mbox{M.\ Karlsson}}
  \altaffiliation[Present address: ]{Nordea Bank Danmark A/S, Christiansbro, \mbox{DK-0900} Copenhagen, Denmark}
\affiliation{Department of Physics, Lund University, SE-221 00 Lund, Sweden}
\author{\mbox{J.-O.\ Adler}}
\affiliation{Department of Physics, Lund University, SE-221 00 Lund, Sweden}
\author{\mbox{L.E.M.\ Andersson}}
\affiliation{Department of Physics, Lund University, SE-221 00 Lund, Sweden}
\author{\mbox{V.\ Avdeichikov}}
\affiliation{Department of Physics, Lund University, SE-221 00 Lund, Sweden}
\author{\mbox{B.L.\ Berman}}
\affiliation{Department of Physics, The George Washington University, Washington DC, 20052, USA}
\author{\mbox{M.J.\ Boland}}
  \altaffiliation[Present address: ]{Australian Synchrotron, Clayton, Victoria 3168, Australia}
\affiliation{MAX-lab, Lund University, SE-221 00 Lund, Sweden}
\author{\mbox{W.J.\ Briscoe}}
\affiliation{Department of Physics, The George Washington University, Washington DC, 20052, USA}
\author{\mbox{J.\ Brudvik}}
\affiliation{MAX-lab, Lund University, SE-221 00 Lund, Sweden}
\author{\mbox{J.R.\ Calarco}}
\affiliation{Department of Physics, University of New Hampshire, Durham, NH 03824, USA}
\author{\mbox{G.\ Feldman}}
\affiliation{Department of Physics, The George Washington University, Washington DC, 20052, USA}
\author{\mbox{K.G.\ Fissum}}
  \altaffiliation{Corresponding author; \texttt{kevin.fissum@nuclear.lu.se}}
\affiliation{Department of Physics, Lund University, SE-221 00 Lund, Sweden}
\author{\mbox{K.\ Hansen}}
\affiliation{MAX-lab, Lund University, SE-221 00 Lund, Sweden}
\author{\mbox{D.L.\ Hornidge}}
  \altaffiliation[Present address: ]{Department of Physics, Mount Allison University, Sackville NB, E4L 1E6 Canada}
\affiliation{Department of Physics and Engineering Physics, University of Saskatchewan, Saskatoon SK, S7N 5E2 Canada}
\author{\mbox{L.\ Isaksson}}
  \altaffiliation[Present address:  ]{MAX-lab, Lund University, SE-221 00 Lund, Sweden}
\affiliation{Department of Physics, Lund University, SE-221 00 Lund, Sweden}
\author{\mbox{N.R.\ Kolb}}
\affiliation{Department of Physics and Engineering Physics, University of Saskatchewan, Saskatoon SK, S7N 5E2 Canada}
\author{\mbox{A.A.\ Kotov}}
  \altaffiliation{deceased}
\affiliation{Petersburg Nuclear Physics Institute, 188350 Gatchina, Russia}
\author{\mbox{P.\ Lilja}}
  \altaffiliation[Present address:  ]{MAX-lab, Lund University, SE-221 00 Lund, Sweden}
\affiliation{Department of Physics, Lund University, SE-221 00 Lund, Sweden}
\author{\mbox{M.\ Lundin}}
  \altaffiliation[Present address:  ]{MAX-lab, Lund University, SE-221 00 Lund, Sweden}
\affiliation{Department of Physics, Lund University, SE-221 00 Lund, Sweden}
\author{\mbox{B.\ Nilsson}}
  \altaffiliation[Present address:  ]{MAX-lab, Lund University, SE-221 00 Lund, Sweden}
\affiliation{Department of Physics, Lund University, SE-221 00 Lund, Sweden}
\author{\mbox{D.\ Nilsson}}
\affiliation{Department of Physics, Lund University, SE-221 00 Lund, Sweden}
\author{\mbox{G.V.\ O'Rielly}}
  \altaffiliation[Present address: ]{Department of Physics, University of Massachusetts Dartmouth, North Dartmouth MA 02747, USA}
\affiliation{Department of Physics, The George Washington University, Washington DC, 20052, USA}
\author{\mbox{G.E.\ Petrov}}
\affiliation{Petersburg Nuclear Physics Institute, 188350 Gatchina, Russia}
\author{\mbox{B.\ Schr\"oder}}
\affiliation{Department of Physics, Lund University, SE-221 00 Lund, Sweden}
\author{\mbox{I.I.\ Strakovsky}}
\affiliation{Department of Physics, The George Washington University, Washington DC, 20052, USA}
\author{\mbox{L.A.\ Vaishnene}}
\affiliation{Petersburg Nuclear Physics Institute, 188350 Gatchina, Russia}

\collaboration{The MAX-lab Nuclear Physics Working Group}
\noaffiliation

\date{\today}

\begin{abstract}

The two-body photodisintegration of $^3$He has been investigated using
tagged photons with energies from 14 -- 31 MeV at MAX-lab in Lund, 
Sweden.  The two-body breakup channel was unambiguously identified by the 
(nonsimultaneous) detection of both protons and deuterons. This approach 
was made feasible by the over-determined kinematic situation afforded by 
the tagged-photon technique.  Proton- and deuteron-energy spectra were 
measured using four silicon surface-barrier detector telescopes located 
at a laboratory angle of $90^\circ$ with respect to the incident photon-beam 
direction.  Average statistical and systematic uncertainties of 5.7\% and 
6.6\% in the differential cross section were obtained for 11 photon-energy 
bins with an average width of 1.2 MeV\@.  The results are compared to 
previous experimental data measured at comparable photon energies as well 
as to the results of two recent Faddeev calculations which employ realistic 
potential models and take into account three-nucleon forces and final-state 
interactions.  Both the accuracy and precision of the present data are 
improved over those obtained in the previous measurements.  The data are in 
good agreement with most of the previous results, and favor the inclusion of 
three-nucleon forces in the calculations.

\end{abstract}


\maketitle

\section{\label{section:introduction}Introduction}

With only three nucleons, $^3$He has long attracted attention as one of the
most straightforward testing grounds for nuclear theories, primarily because
numerically accurate solutions to this quantum-mechanical three-body problem 
exist.  Further, when low-energy photons are used as a probe, these 
calculations can be carried out in a non-relativistic framework and with a 
well-understood initial-state interaction \cite{Ber64}.

Photodisintegration experiments on $^3$He have been performed 
for a long time, and over the years several experiments at low photon energies 
have been reported \protect\cite{Ber64,Ste65,Kun71,Tic73,Cha74,Sko79,Fau81} 
(see for example Fig. \protect 1 in Ref.\ \protect\cite{Sko79} where the 
previous results are presented in the c.m. system and our 
Fig.\ \protect\ref{figure:previous_data} where the previous results are 
presented in the lab system.  At these energies, the differences are small -- 
on the order of 1--2\%).  The general trend in the results is clear: the 
differential cross-section data obtained at a lab angle of 90$^{\circ}$ as a 
function of photon energy ($E_\gamma$) rises sharply from the reaction 
threshold at 5.49 MeV to a peak value which occurs at around 11.5 MeV\@.  Above 
12 MeV, the cross section decreases smoothly.  Beyond the general trend, a 
clear question arises regarding the normalizations of the various data sets. 
Around 12 MeV, the measured cross-section data range from 90 to 120 $\mu$b/sr. 
At higher energies, fewer data sets exist, but they appear to converge. If the 
previous measurements are investigated in more detail, another feature 
appears -- the 
cross section determined in experiments performed with untagged bremsstrahlung 
photons \protect\cite{Ber64,Ste65,Tic73} tends to be smaller than the cross 
section determined in either radiative-capture experiments \protect\cite{Sko79}
or electrodisintegration experiments \protect\cite{Kun71,Cha74}.

Recent advances in theoretical techniques and computational capabilities have 
made it possible to perform complete Faddeev calculations for the $A=3$ 
system \protect\cite{Ski03,Del05}. These calculations use modern, realistic 
potentials taking into account three-nucleon forces (3NF) as well as
final-state interactions (FSI).  In particular, the calculations below 
$E_\gamma = 20$ MeV are sensitive to the choice of nucleon-nucleon (NN) 
potential, 3NF effects, and Coulomb effects. However, the quality of the 
existing data has made comparisons between theoretical predictions and 
experimental results difficult. The need for more accurate and precise 
measurements is clear -- especially below 20 MeV where sensitivity is high and 
experimental discrepancies are most pronounced.

A recent major improvement in experimental technique has been the development
of accelerators capable of providing almost continuous electron beams which
have facilitated the method of photon tagging.  With this method, it is 
possible to directly measure both the energy of the photon as well as the total
number of photons incident upon the target.  The tagged-photon technique also 
facilitates the discrimination between two-body and three-body 
photodisintegration via kinematic overdetermination.  This means that in a 
modern tagged-photon experiment on $^{3}$He, two-body photodisintegration 
events can be identified by the detection of the proton only, substantially 
less challenging than the detection of a deuteron.

In this article, we present a comprehensive new data set for the 
photodisintegration of $^{3}$He near threshold obtained using tagged photons 
with energies from \mbox{14 -- 31 MeV}\@.  We compare our data with previous 
measurements as well as modern calculations \cite{Ski03,Del05} which employ 
the Faddeev technique and include final-state interactions.  A detailed 
description of the experiment is presented in Ref.\ \protect\cite{kar05}.

\section{\label{section:expt}Experiment}

\begin{figure}[!h]
\begin{center}
\resizebox{0.5\textwidth}{!}{\includegraphics{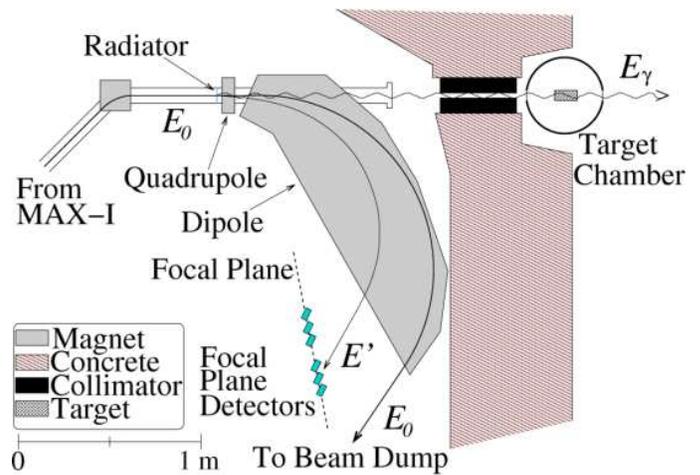}}
\caption{(Color online) An overview of the experimental layout at the time of 
the experiment.  The beam passed through a radiator generating bremsstrahlung.
Non-interacting electrons passed to a well-shielded beam dump.  Recoil 
electrons were momentum-analyzed using a magnetic tagging spectrometer equipped
with a 64-counter focal-plane scintillator array.  The resulting tagged-photon 
beam was collimated before it entered the target chamber.  Reaction products 
were detected using 4 silicon-detector telescopes mounted at 90$^{\circ}$ to 
the direction of the photon beam (see 
Fig.\ \ref{figure:target_detector_setup}).  See text for further details.}
\label{figure:exp_schematic}
\end{center}
\end{figure}

The experiment was performed at the tagged-photon facility \cite{Adl97} located
at MAX-lab \cite{maxlab}, in Lund, Sweden.  A pulse-stretched electron beam 
with a nominal energy of 93 MeV, a nominal current of 20 nA, and a nominal duty 
factor of 75\% was used to produce quasi-monochromatic photons via the
bremsstrahlung-tagging technique \cite{Adl90}.  A diagram of the experimental
layout is shown in Fig.\ \ref{figure:exp_schematic}.

\subsection{\label{subsection:beam}Photon beam}

A 0.1\% radiation-length Al radiator was used to generate a 
bremsstrahlung photon beam from the electron beam.  Non-radiating electrons 
passed to a well-shielded beam dump (see Fig.\ \ref{figure:exp_schematic}).  
Post-bremsstrahlung electrons were momentum-analyzed using a magnetic 
spectrometer equipped with a 64-counter focal-plane scintillator array.  The 
nominal photon-energy resolution of 270 keV resulted almost entirely from the 
6.2 mm width of a single focal-plane counter.  

The scintillators were mounted in two modules consisting of 32 non-overlapping 
counters, and the tagged photon-energy range (14 -- 31 MeV) was selected by 
sliding the array to the appropriate position along the focal plane of the 
spectrometer.  The analog signals from each of the focal-plane detectors were 
individually discriminated and then passed to scalers and TDCs.  The average 
single-counter rate during these measurements was 2 MHz per MeV\@.

The size of the photon beam was defined by a tapered tungsten-alloy primary 
collimator of 12.3 mm diameter.  The primary collimator was followed by a 
dipole magnet and a post-collimator which were used to remove any charged 
particles produced in the primary collimator.  The position of the photon beam 
both upstream and downstream of the target location was determined by 
irradiating Polaroid film after every adjustment of the electron beam.  In this
manner, the beam spot was determined to be 18.6 $\pm$ 0.4 mm in diameter
throughout the target cell.  As a uniform distribution of events across the
detectors was observed throughout the experiment, the dislocation of the
symmetry axis of the target chamber with respect to the direction of the
photon beam was determined to be less than 2 mm (see 
Sect. \ref{subsubsection:acceptance}). 

The tagging efficiency \cite{Adl97} is the ratio of the number of tagged
photons which struck the target to the number of recoil electrons which were
registered by the associated focal-plane counter.  It was measured absolutely 
(using a 100\% efficient lead/scintillating-fiber photon detector) on a 
frequent basis during the experiment.  These measurements required a very low 
intensity photon beam to avoid pileup in the photon detector, and were 
corrected for accelerator-associated (such as scattering of recoil electrons 
between focal-plane detectors), radiator-associated (such as multiple Coulomb 
and M\"{o}ller scattering) and room-associated (such as activation) sources of 
background.  Tagging efficiency was typically 18\%, resulting from the strict 
collimation of the photon beam.  Uncertainty in
the tagging efficiency had both scale and rate-dependent contributions.  
The scale contributions included differences in the duty factor between tagging
efficiency and production runs (2\%), time-dependent variations (2\%), and a 
focal-plane detector live time correction (3\%).  The rate-dependent component 
arose from the sensitivity of the tagging efficiency to background in the 
focal-plane detectors.  This background arose from the post-radiator, enlarged,
unconverted primary electron beam interacting with the pole surfaces of the 
tagging spectrometer.  Focal-plane detectors corresponding to the lower photon 
energies were located closer to this electron beam, and thus suffered a much 
higher rate of background events.  The corresponding average uncertainty 
associated with the rate-dependent correction was 5\%.

\subsection{\label{subsection:target}Target}

\begin{figure}[!h]
\begin{center}
\resizebox{0.45\textwidth}{!}{\includegraphics{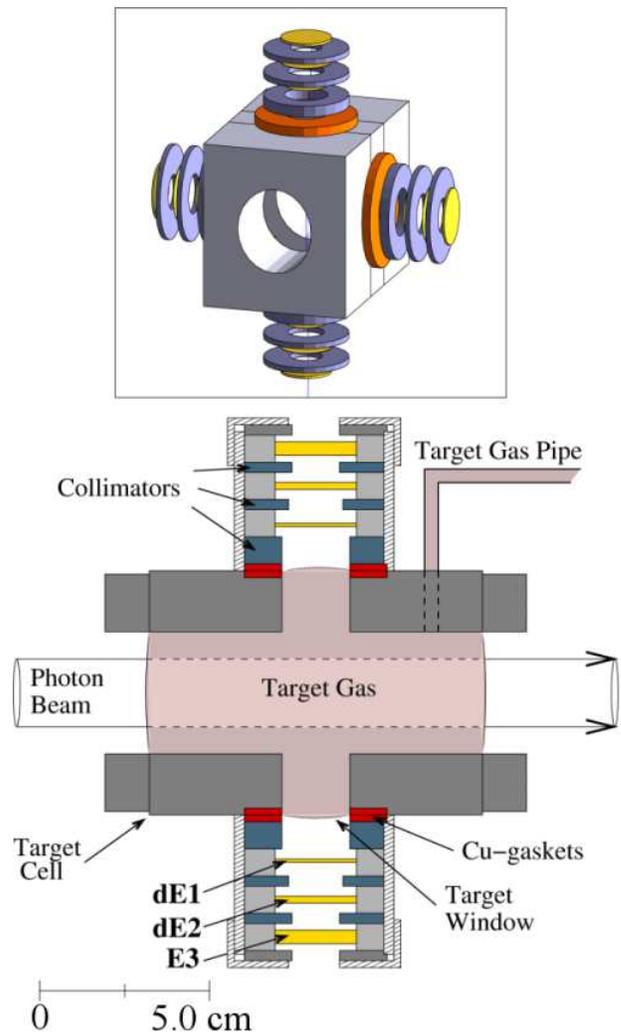}}
\caption{(Color online) The top panel shows a three-dimensional representation 
of the target/detector setup.  The bottom panel shows a schematic of the target
cell.  Two of the four detector telescopes are shown together with the 
photon-beam envelope and the target-gas volume.}
\label{figure:target_detector_setup}
\end{center}
\end{figure}

The target cell is presented in Fig.\ \ref{figure:target_detector_setup}.  It 
was manufactured from a single block of stainless steel with dimensions 
\mbox{70 $\times$ 70 $\times$ 100 mm$^3$}. One large circular hole was milled 
through the block creating the entrance and exit ports for the photon beam.  
Perpendicular to this hole, two smaller holes were milled through the block, 
creating four exit ports for the reaction products.  These four exit ports were
``aimed" at the center of the gas volume.  The target cell was sufficiently 
robust that it also served as the base for mounting the detector systems 
installed at each detector port.

\color{black}
Thin havar foils were used to seal all of the cell ports and thus contain the
gaseous $^3$He that served as the target.  The beam entrance 
and exit windows were 12.5 $\mu$m thick.  During the experiment, the 
detector-port windows were reduced to 5 $\mu$m in thickness.  
The foils were mounted in position by clamping them between two copper gaskets 
together with an indium seal.  This resulted in a very low rate of target-gas 
leakage.  The target cell was placed in a vacuum chamber maintained
at $5\times 10^{-3}$ torr.  This target chamber was placed as close as possible
(the entrance window to the chamber was approximately 10 cm downstream of the
exit aperture of the photon-beam collimator) to the photon-beam collimator to 
ensure that the photon-beam envelope was completely subtended by the target-gas 
volume.

The target-cell filling system consisted of a vacuum pump, a precision pressure 
gauge, and a gas cylinder.  The cell was filled with 99.96\% pure $^3$He 
gas to a nominal pressure of 2 bar.  The target pressure was continuously 
monitored throughout the experiment.  The pressure change over the course of a 
two-week run period was typically 2\% so that additional filling of the target 
cell during a run period was unnecessary.  The effective target pressure was 
taken to be the pressure in the middle of the run period.

The target density was determined based upon the assumption that the target gas
behaved as an ideal gas. The temperature used in determining the target 
thickness was taken to be the room temperature measured close to the target 
chamber. Typical values were about 25.5 $^\circ$C with a systematic uncertainty
of less than 0.5\%.  The variation in the target pressure with time together
with the accuracy of the pressure gauge used to monitor the system resulted in
a total systematic uncertainty in the target density of 2\%.

Empty-target measurements were performed with the target-cell pressure at 0.1 
torr.  It was determined that this background contribution to the data was
negligible.

\subsection{\label{subsections:ssbs}Detector telescopes}

Protons and deuterons were detected in four detector telescopes.  Each 
telescope consisted of three totally depleted silicon surface barrier 
detectors mounted in a $\Delta$E--$\Delta$E--E (labeled dE1--dE2--E3 in 
Fig.\ \ref{figure:target_detector_setup}) configuration. Silicon detectors have 
essentially 100\% intrinsic detection efficiency for charged particles.  For 
each of the the dE1 detectors, two detector thicknesses were used: 25 $\mu$m 
and 50 $\mu$m.  The dE2 and E3 detectors were 150 $\mu$m and 1000 $\mu$m thick, 
respectively.  This arrangement, in which the dE2 detector could serve as 
either a $\Delta$E or E detector, increased the energy range over which 
particle identification (PID) could be performed. 

A collimator was located in front of each detector element. For protons or
deuterons originating from an interaction at the center of the target cell, 
these collimators defined the point-source solid angle subtended by the 
detectors.  These point-source solid angles were 107, 68, and 38 msr for the 
dE1, dE2, and E3 detectors respectively.  For points of interaction further 
away from the center of the target cell, the copper gaskets as well as the 
target cell itself contributed, creating a complex geometric acceptance. The 
determination of this geometric acceptance was performed using a 
\textsc{geant4} \cite{Ago03} Monte-Carlo simulation, which accounted for the 
uniform photon-beam profile of 18.6 mm diameter, the 6 cm extended-target 
geometry, and the collimator arrangement in front of each detector element.

\subsection{\label{subsection:acq}Electronics and data acquisition}

\begin{figure}[!h]
\begin{center}
\resizebox{.45\textwidth}{!}{\includegraphics{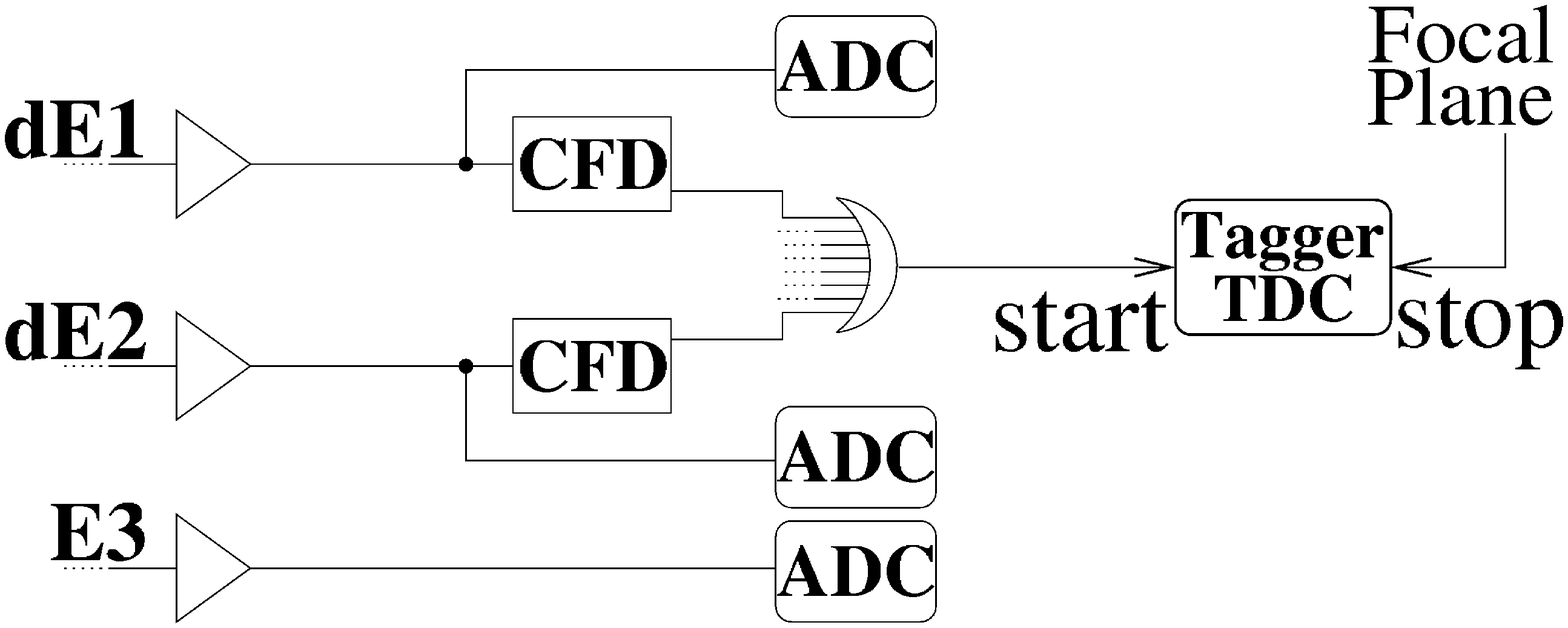}}
\caption{{Schematic diagram of the electronics used for a single telescope
together with the focal plane.}}
\label{figure:electronics}
\end{center}
\end{figure}

An overview of the electronics is shown in Fig.\ \ref{figure:electronics}.

The analog signal from each silicon detector was sent to a preamplifier.
The preamplifiers were located approximately 30 cm from the target 
chamber. The output signals from the preamplifiers were then passed to
timing-filter amplifiers (TFA). The output from a given TFA was 
symmetrically divided for the dE1 and dE2 detectors, and passed both to an 
analog-to-digital converter (ADC) and a constant-fraction 
discriminator (CFD). The CFD determined the timing of the dE1 and dE2 event 
triggers. Output signals from the CFDs were also used to stop TDCs.
The TFA output for the E3 detector was simply passed to an ADC. For the thick 
dE1, dE2, and E3 detectors, the ADC used was charge-integrating.
For the thin dE1 detectors, a spectroscopy amplifier and a
peak-sensing ADC were used together with modified preamplifiers to
allow for reasonable particle identification when the deposited energy was 
relatively small and the noise in the detectors was relatively large.

The OR of the CFD signals from the eight $\Delta$E (4 dE1 and 4 dE2) detectors
was used to generate the trigger signal.  The trigger signal was used to start 
the focal-plane TDCs, gate the ADCs, and start TDCs connected to 
the $\Delta$E detectors to determine the relative timing between them. This 
was necessary to account for the different charge-collection times of the 
different thickness silicon detectors.  Note that the dE2 detectors had the 
best time resolution so that they were used to determine the overall trigger 
timing.

The trigger signal was also passed to a VME front-end computer to initiate the 
data readout from the CAMAC crates via a branch driver and an inhouse software 
toolkit.  Resulting events were defined and analyzed offline using 
the \textsc{root} toolkit \cite{root}.

\section{\label{section:data_analysis}Data Analysis}

\subsection{\label{subsection:energy_calibration}Calibration of the silicon detectors}

\begin{figure}[!h]
\begin{center}
\resizebox{.5\textwidth}{!}{\includegraphics{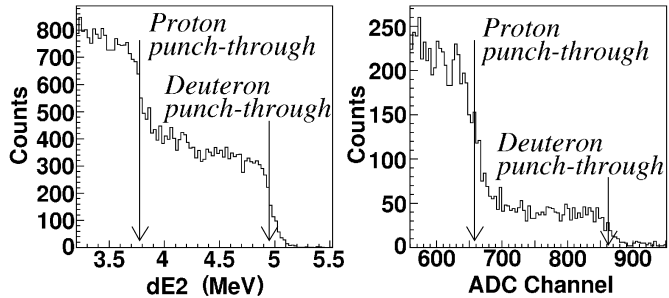}}
\caption{{An illustration of the energy-calibration procedure for the dE2 
detector. The punch-through energies are indicated by the arrows.  The left 
panel shows the simulated energy spectrum while the right panel shows the 
corresponding measured ADC spectrum.  See text for details.}}
\label{figure:dE2energycal}
\end{center}
\end{figure}

The calibration of the dE1 and dE2 detectors was straightforward as the 
thicknesses of the detectors were specified to within 2\%, equivalent to the
precision of the \textsc{geant4} energy-loss corrections.
Thus, deuterons (or protons) had a specific energy required for a particle to 
barely pass through a detector of a given thickness (the ``punch-through" 
energy). ADC spectra were compared to \textsc{geant4}-simulated spectra to
determine the punch-through energy (see Fig.\ \ref{figure:dE2energycal}).  

For each of the dE1 and dE2 detectors, the correspondence (hereafter refered to
as the ``gain") between the simulated punch-through energy (in MeV) and the 
measured punch-through energy (in ADC channels) was given by

\begin{equation}
g=\frac{E_{\rm punch-through}}{ADC_{\rm punch-through}},
\label{equation:gain}
\end{equation}
where $E_{\rm punch-through}$ was the punch-through energy from the simulated 
spectrum and $ADC_{\rm punch-through}$ was the corresponding punch-through 
position in the ADC spectrum.  Each detector thus yielded two punch-through 
calibration points: one for protons and one for deuterons. The resulting gains 
determined from the two particle types were equal to better than 2\%. The final
uncertainty in the gain factors was dominated by the uncertainty in detector 
thicknesses and was estimated to be 3\%.

The calibration scheme for the dE1 and dE2 detectors was not applicable to the 
E3 (1000 $\mu$m) detectors. This was because the energy deposited for the 
punch-through events was large enough to saturate the amplifiers.  Instead,
coincidences between dE2-deuterons and E3-protons in opposed telescopes were 
examined.  

For a given E3-proton or dE2-deuteron, the detected energy $T'_{p,d}$ was given
by $T'_{p,d}=T_{p,d}-L_{p,d}$, where $T_{p,d}$ was the kinetic energy of the 
E3-proton or dE2-deuteron in question and $L_{p,d}$ was its energy loss.  By 
differentiating the expressions for E3-protons and dE2-deuterons and then 
dividing the resulting equations, the following expression was obtained:

\begin{equation}
\frac{dT^{\prime}_{p}}{dT^{\prime}_{d}} = 
\frac{dT_{p}}{dT_{d}} \cdot \frac{1-\frac{dL_{p}}{dT_{p}}}{1-\frac{dL_{d}}{dT_{d}}}.
\label{equation:dTddTp}
\end{equation}
In this expression, the quotient $\frac{dT_{p}}{dT_{d}}$ is a kinematic factor 
that is completely dominated by the relative masses of the proton and deuteron.
For energies well away from reaction threshold, it is almost invariant with 
respect to photon energy.  While the loss functions are strongly dependent upon
energy, their ratio is not. 

\begin{figure}[!h]
\begin{center}
\resizebox{.5\textwidth}{!}{\includegraphics{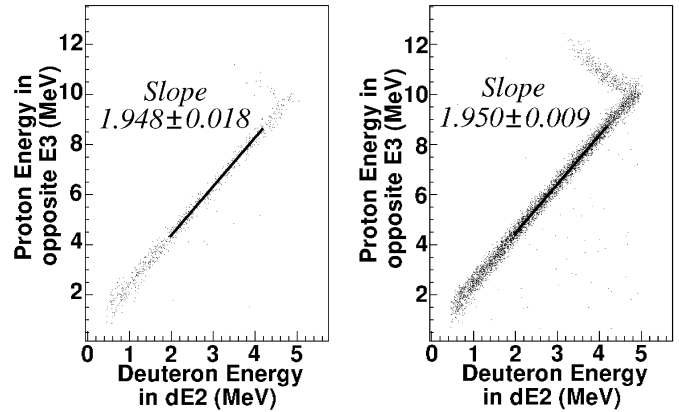}}
\caption{An illustration of the calibration of an E3 detector.  Each panel
shows E3-protons plotted against dE2-deuterons.  The left panel shows 
\textsc{geant4}-simulated events for an E3 gain which resulted in the slopes 
in the distributions being equal, while the right panel shows data.  When the 
slopes in the two scatter plots were equal, the gain in the E3 detector had 
been determined.  See text for details.}
\label{figure:E30PvdE2}
\end{center}
\end{figure}

Figure \ref{figure:E30PvdE2} shows E3-proton energy plotted against
dE2-deuteron energy \textsc{geant4}-simulated events (left panel) and for data 
(right panel).  Coincident events lie in a well-defined linear band. The slope
of this band is related to the ratio of the gains in the detectors (recall
Eq. \ref{equation:dTddTp}).  Using the gain in the dE2 detector determined 
using the punch-through method, data were simulated using \textsc{geant4} 
varying the gain in the E3 detector until the slope in the simulated 
scatter plot was equal to the slope in the data scatter plot.  When the slopes
in the two scatter plots were equal, the gain in the E3 detector had been 
determined.

A distinct advantage to performing the energy calibration of the E3 detector
using this method is that it is not as sensitive to uncertainties in the
energy-loss functions or detector geometries and thicknesses.  The uncertainty 
in $\frac{dT^{\prime}_{p}}{dT^{\prime}_{d}}$ was determined to be \mbox{1\%}.

Uncertainties in the energy-calibration procedure resulted in a systematic 
uncertainty in the knowledge of the photon energies corresponding to each 
tagger channel of about 200 keV.  This contributed a 1.5\% uncertainty to the 
measured cross section.

\subsection{\label{subsection:pid}Particle identification (PID)}

\begin{figure}[!h]
\begin{center}
\resizebox{.5\textwidth}{!}{\includegraphics{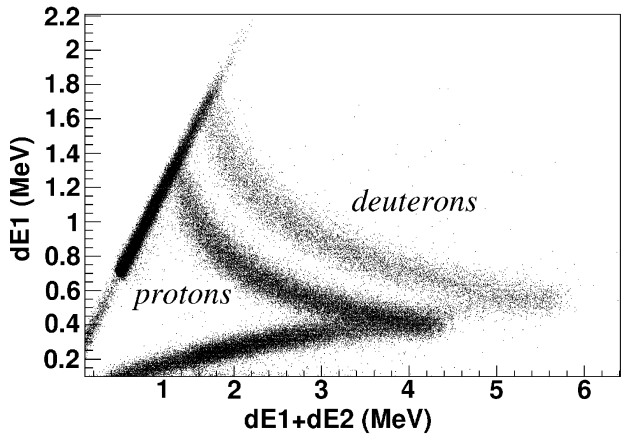}}
\caption{An energy-loss scatter plot showing the detector-element combination 
dE1 versus (dE1+dE2).  Both the proton band and the deuteron band may be easily 
identified.  Higher energy ``punch-through" or "back-bending" events were 
analyzed in more detail in dE2 versus (dE2+E3) scatter plots.  See text for 
details.}
\label{figure:dE1vdEtot}
\end{center}
\end{figure}

Charged reaction products were identified by their energy loss in 
the $\Delta$E detectors.  Because the detector telescopes used for this 
experiment each consisted of three detector elements, two different 
combinations of elements could be used for energy-based PID.  For 
low-energy particles, dE1 versus (dE1+dE2) scatter plots were filled; and for 
high-energy particles, dE2 versus (dE2+E3) scatter plots were filled. This 
provided high-quality particle identification over the entire range of particle
energies available to this experiment.  A typical scatter plot is shown in 
Fig.\ \ref{figure:dE1vdEtot}, where a clear separation between protons and 
deuterons may be observed. The detection efficiency for both protons and 
deuterons in the silicon detectors was 100\%.  

\subsection{\label{subsection:background_subtraction}Background subtraction}

\begin{figure}[!h]
\begin{center}
\resizebox{.5\textwidth}{!}{\includegraphics{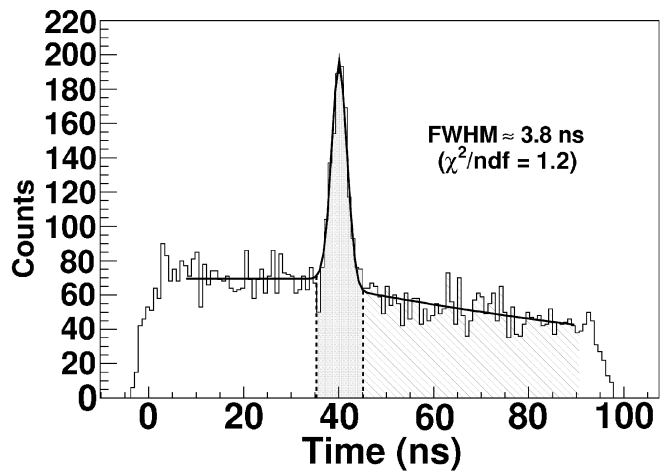}}
\caption{A TDC spectrum corresponding to a photon-energy bin of 
3.25 MeV demonstrating coincidences between post-bremsstrahlung electrons in 
the tagger focal plane and charged particles in the silicon detectors.  This 
spectrum has been subjected to a missing energy cut to improve the 
signal-to-noise ratio.  A FWHM timing resolution of 3.8 ns was obtained.  See 
text for details.} 
\label{figure:TDCspectrum_FP_20_31}
\end{center}
\end{figure}

Timing information from the tagger was used to establish coincidences between 
the protons and deuterons in the silicon detectors and post-bremsstrahlung 
electrons 
in the focal-plane detectors.  Fig. \ref{figure:TDCspectrum_FP_20_31} shows a
typical timing distribution for a 3.25 MeV wide photon-energy bin.  In this 
distribution, the shaded prompt region contains an easily identified 
coincidence peak superimposed on top of a random background.  The function 
fitted to this spectrum consisted of Gaussian superimposed upon a constant 
random background to the left of the timing peak (the timing region where 
``stolen coincidences" \cite{Owe90,Hor03} may occur -- see below) and a 
decaying exponential to the right of the timing peak (the cross-hatched timing 
region where the coincidences are truly random).  A FWHM-timing resolution of 
3.8 ns was obtained.  In order to separate the true coincidences from the 
random coincidences, a cut (indicated by the vertical bars) was placed on the 
prompt peak and to the right of the prompt peak (indicated by the fitted 
region) and two ``missing-energy" spectra were filled.

\begin{figure}[!h]
\begin{center}
\resizebox{.5\textwidth}{!}{\includegraphics{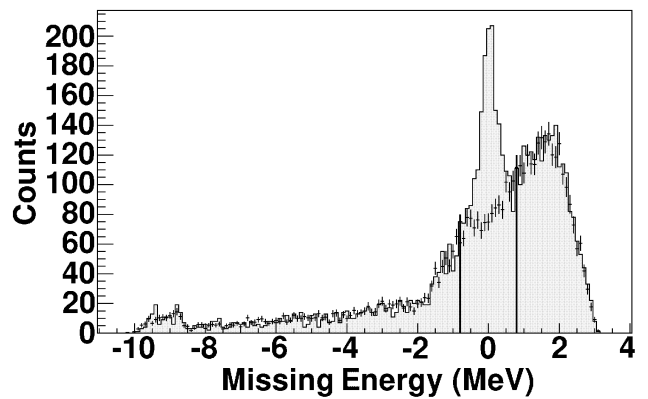}}
\caption{{A prompt (both true tagged and random events) 
$E_{\rm miss}$ spectrum together with a normalized purely random $E_{\rm miss}$
spectrum. The vertical bars indicate the region where true tagged events 
appeared.  See text for details.}}
\label{figure:Emiss_PTnormRD_FP_20_31}
\end{center}
\end{figure}

The missing energy $E_{\rm miss}$ was defined as
\begin{equation}
\label{equation:missingenergy}
E_{\rm miss}=T_d(E_\gamma) - L_d(E_\gamma) - E_{\rm total},
\end{equation}
where $T_d(E_\gamma)$ was the kinetic energy of the deuteron at the reaction 
vertex and $L_d(E_\gamma)$ was the energy loss experienced in the gas and in 
the target-window foils by the deuteron.  $E_{\rm total}$ was the (total) 
detected energy in a given telescope. Note that the term $T_d(E_\gamma) - 
L_d(E_\gamma)$ was not only a function of photon energy but also of the 
deuteron emission angle.  The \textsc{geant4} simulation was used to determine 
the average value of this term for each focal-plane detector.  In this manner, 
true tagged deuteron events were localized to a peak at \mbox{$E_{\rm miss}$ 
= 0 MeV}, whereas random events populated a much larger range. Figure 
\ref{figure:Emiss_PTnormRD_FP_20_31} shows the prompt and normalized random 
$E_{\rm miss}$ spectra corresponding to the TDC spectrum shown in Fig.\ 
\ref{figure:TDCspectrum_FP_20_31}.  Based upon the assumption that the shape of
the $E_{\rm miss}$ spectrum for random events within the prompt region was the 
same as that for events from the purely random region, the normalization was 
determined by requiring the number of events outside the peak region to be the 
same in the prompt $E_{\rm miss}$ spectrum and the normalized random 
$E_{\rm miss}$ spectrum.

\begin{figure}[!h]
\begin{center}
\resizebox{.5\textwidth}{!}{\includegraphics{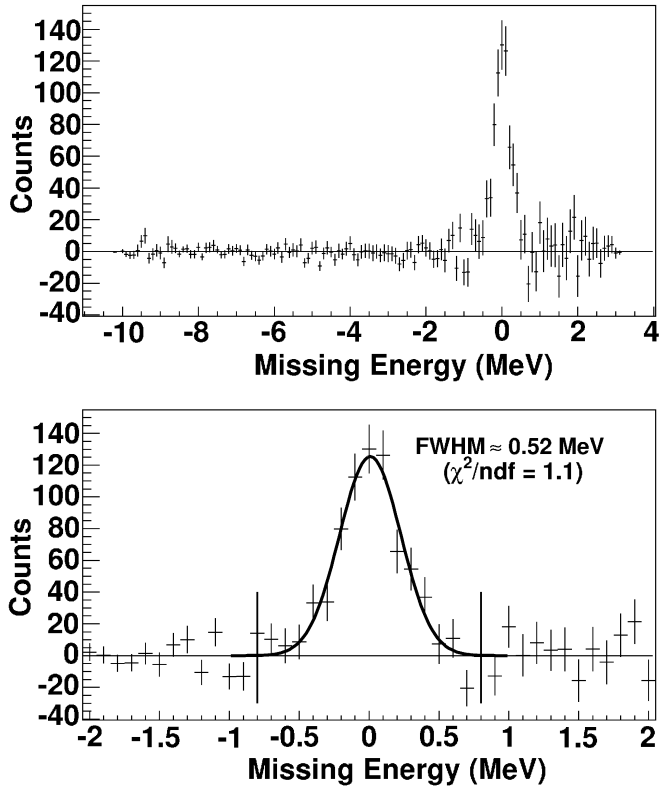}}
\caption{{Top panel:  the difference between the prompt and normalized 
random spectra shown in Fig.\ \ref{figure:Emiss_PTnormRD_FP_20_31}.  Away from
the clearly visible peak, the data are structureless and statistically 
consistent with zero.  Bottom panel:  a Gaussian distribution has been fitted
to the peak to determine the FWHM energy resolution as well as the region over 
which the peak data were integrated.  See text for details.}}
\label{figure:Emiss_sub_FP_20_31}
\end{center}
\end{figure}

The top panel of Fig.\ \ref{figure:Emiss_sub_FP_20_31} shows the difference 
between the prompt and normalized random spectra shown in Fig.\ 
\ref{figure:Emiss_PTnormRD_FP_20_31}.  Away from the peak region at
$E_{\rm miss} = 0$ MeV, the data are flat as a function of energy and 
statistically consistent with zero.  This confirms the validity of the method 
used for the background subtraction.  In the bottom panel of Fig.\ 
\ref{figure:Emiss_sub_FP_20_31}, a Gaussian distribution was fitted to the data
in the vicinity of $E_{\rm miss} = 0$ MeV\@.  The fitted function provided a
constraint on the energy region over which the data were summed bin by bin to 
determine the measured yield.

Stolen coincidences occurred when an uncorrelated (random) post-bremsstrahlung 
electron stopped the focal-plane TDCs prior to a (true) post-bremsstrahlung 
electron correlated in time with a charged-particle event. A correction was 
applied to account for these events, which would otherwise have been missed. 
Due to the very high event rates observed in this experiment (from 2 to 5 MHz),
this correction to the yield ranged from 6\% to 28\%.  A detailed discussion
of the correction is presented in Ref.\ \cite{kar05}.

An identical analysis was performed on the empty-target data and demonstrated 
that there was no measurable contribution to the full-target spectra.

\subsection{\label{subsection:cross_section}Cross Section}

The laboratory differential cross section for the reaction
\mbox{$\gamma$ $+$ $^3$He $\rightarrow$ d + p} for each photon energy bin was 
extracted using

\begin{equation}
\frac{d \sigma}{d \Omega} (E_{\gamma}) = 
\frac{Y_{d,p}(E_{\gamma})}
{N_\gamma(E_{\gamma}) \cdot \rho \cdot \Delta\Omega \cdot l},
\label{equation:cross_section}
\end{equation}
where $Y_{d,p}(E_{\gamma})$ was either the true deuteron or true proton yield 
corrected for stolen coincidences and electronic deadtime effects; 
$N_\gamma(E_{\gamma})$ was the total number of photons for a given 
photon-energy bin given by the product of the tagging efficiency and the 
number of electrons registered in the corresponding focal-plane scalers and 
corrected for electronic deadtime effects; $\rho$ was the target density 
(recall Sec. \ref{subsection:target}); and $\Delta\Omega$ was the effective 
solid angle for an extended target of length $l$.  The cross section was 
evaluated separately for each focal-plane detector and subsequently binned.

\subsubsection{\label{subsubsection:yield}Yield}

\begin{figure}[!h]
\begin{center}
\resizebox{.5\textwidth}{!}{\includegraphics{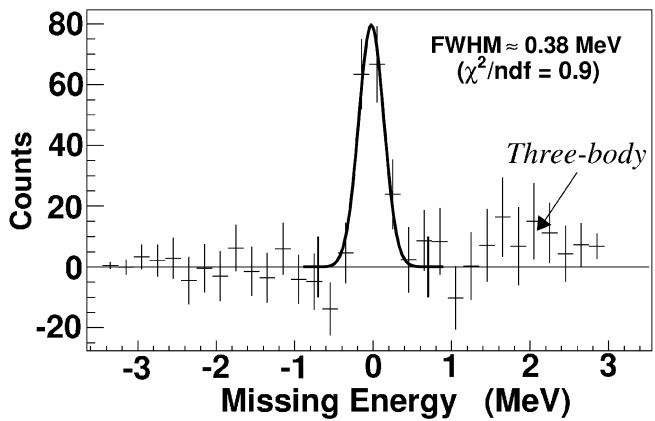}}
\caption{{An $E_{\rm miss}$ spectrum for events identified as protons.  
2bbu protons are clearly confined to a peak at zero missing energy, whereas 
3bbu channel protons necessarily have \mbox{$E_{\rm miss}$ $>$ 1.5 MeV}\@.  The 
energy resolution was thus sufficient to separate 2bbu protons from 3bbu 
protons and thus allow the cross section to be investigated in a complementary
manner via the 2bbu proton events.}}
 \label{figure:emiss_proton}
\end{center}
\end{figure}

The yield of true deuteron events corrected for electronic deadtime effects 
was used to determine the cross section.  However, in a measurement of the 
\mbox{$\gamma$ $+$ $^3$He $\rightarrow$ d + p} reaction 
where the photon energy is known, a measurement of the kinetic energy of the 
proton with a sufficiently high energy resolution together with knowledge of 
the proton 
angle allows for unambiguous identification of the two-body breakup (2bbu) 
channel.  As this work was performed using tagged photons and the energy 
resolution was sufficiently high (approximately 0.5 MeV, see 
Fig.\ \ref{figure:emiss_proton}), the over-determined kinematic situation arose,
and protons from the photodisintegration of $^3$He were also used to identify 
2bbu events \footnote{
The differential cross section calculated using the proton yield was thus
measured at $\theta_p=90^\circ$.  At the photon energies employed here, this 
corresponds to the deuteron being ejected at about $\theta_d \approx 82^\circ$.
Angular distributions provided by Schadow {\it et al.} \cite{Sch01,Sch01pc}
were thus employed to transform the cross section measured at 
$\theta_p=90^\circ$ to the cross section at $\theta_d = 90^\circ$ using the 
relation
\mbox{
$\frac{d\sigma}{d\Omega}\left(\theta_d = 90^\circ\right) = 
\frac{d\sigma}{d\Omega}\left(\theta_p=90^\circ\right)\cdot\frac{w[\theta_d=90^\circ]}{w[\theta_d(\theta_p=90^\circ)]}$},
where the function $\theta_d(\theta_p=90^\circ)$ was the angle of the deuteron 
corresponding to a proton angle of $90^\circ$. Typical values of the factor 
$w[\theta_d=90^\circ] / w[\theta_d(\theta_p=90^\circ)]$ were in the range   
1.03 to 1.10.
}.

In the final cross-section results, events from as many as twelve focal-plane 
detectors were combined into a single photon-energy bin, taking into account 
both the variation in the tagged-photon flux as well as the energy-dependence 
of the cross section.  The systematic uncertainty associated with particle
identification and yield determination was $<$1\%.

\subsubsection{\label{subsubsection:no_photons}Number of photons}

The procedure used for obtaining the number of photons incident on the target 
is presented in detail in Ref.\ \cite{Adl97}.  The incident photon flux for each
photon-energy bin was determined by counting the number of recoil electrons in 
the tagger focal plane and correcting the result for electronic deadtime and
the measured tagging efficiency, which was on average 18\%.  Attenuation of 
photon flux due to atomic processes 
within a 9 cm diameter liquid $^{4}$He target was thoroughly investigated and
shown to be negligible in Ref.\ \cite{Nil07}.  It was thus also concluded to be 
negligible for the 6 cm long gaseous $^{3}$He target used in this experiment.

\subsubsection{\label{subsubsection:acceptance}Acceptance}

Due to the use of an extended target, no 
simple analytical expression for the solid angle subtended by the silicon 
detectors could be employed. Instead, the \textsc{geant4} Monte-Carlo 
simulation of the setup was used to quantify the acceptance.

\begin{figure}[!h]
\begin{center}
\resizebox{0.5\textwidth}{!}{\includegraphics{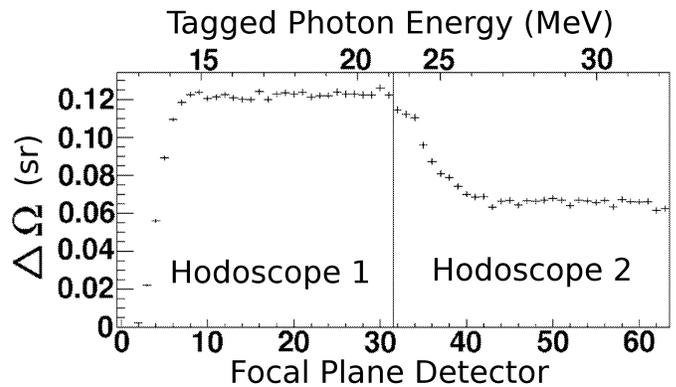}}
\caption{The effective solid angle $\Delta\Omega$ as a function of focal-plane 
detector (photon energy) evaluated for the 6 cm extended target thickness. The 
steep increase in effective solid angle up to 15 MeV was due to threshold 
effects in the PID process. The decrease in effective solid angle which occurs 
between 20 and 25 MeV was a consequence of the collimators located between the 
dE2 and E3 silicon detectors.}
\label{figure:effsolidangle}
\end{center}
\end{figure}

The geometrical acceptance as a function of photon energy was evaluated by 
examining the ratio of the number of detected events ($N_{\rm detected}$) to
the number of generated events ($N_{\rm generated}$).  The number of detected
events was determined by applying the same cuts to the pseudodata that were
used for the analysis of real data.  In this manner, the effective solid angle 
was given by 
\begin{equation}
\Delta\Omega=\frac{ 4\,\pi \cdot N_{\rm detected}}{N_{\rm generated} \cdot w(\theta=90^\circ)},
\end{equation}
where $w(\theta=90^\circ)$ were the LAB angular distributions at LAB angle 
$\theta$ provided by Schadow {\it et al.} (see Sec. \ref{section:results}). 
The effective solid angle as a function of focal-plane detector (photon energy)
evaluated for the 6 cm extended target thickness is shown in 
Fig.\ \ref{figure:effsolidangle}.

The systematic uncertainty in the effective solid angle was dominated by the
positioning of the target chamber and the uncertainty in the locations of the
collimators.  Simulation demonstrated that an 8 mm mispositioning of the target 
chamber (and thus an offset of the photon-beam trajectory with respect to the 
locations of the detector telescopes) introduced a significant skewing of the
yields in the four telescopes together with only a 3.5\% overall reduction.  
Since this skewing of the yields was not observed, it was concluded that the 
target chamber was positioned correctly.  Simulation also demonstrated that the
uncertainty in the locations of the collimators resulted in an uncertainty in
the effective solid angle of 2\%.

\subsubsection{Systematic uncertainties} \label{subsubsection:systematic_uncertainities}

The systematic uncertainty in the measurement was dominated by the systematic
uncertainty in the determination of the number of photons, which ranged from
4\% at $E_{\gamma}$ $=$ 31.2 MeV to 14\% at $E_{\gamma}$ $=$ 14.0 MeV\@.  A 
summary of the systematic uncertainties is presented in Table 
\ref{table:systematic_uncertainties}.  The systematic uncertainties associated 
with each of the individual cross-section data points are presented in Table 
\ref{table:data}. See also the uncertainty bands shown in Figs.\ 
\ref{figure:previous_data} and \ref{figure:calculations}.

The systematic uncertainty associated with the target cell and detector 
telescopes used in this experiment was carefully studied in a previous 
measurement of the tagged two-body photodisintegration of $^3$He performed
at SAL \cite{Ori09} by measuring the $^2$H($\gamma,p)$n total cross section
in \mbox{1 MeV} bins from \mbox{18 $<$ $E_{\gamma}$ $<$ 39 MeV} using the 
exact same target cell and detector telescopes.  Agreement between these SAL 
data and those of Bernabei {\it{et al.}} \cite{Ber86} was excellent.  Further, 
as the agreement between our $^3$He data and these SAL $^3$He data is also 
excellent, we conclude that we have a very good understanding of our systematic 
uncertainties.

\begin{table}
\caption{\label{table:systematic_uncertainties}
A summary of the systematic uncertainties in the cross-section data.}
\begin{ruledtabular}
\begin{tabular}{rcc}
                                 quantity &            uncertainty \\
\hline
               tagging efficiency (scale) &                    4\% \\
tagging efficiency ($<$rate dependent$>$) &                    5\% \\
                   geometrical acceptance &                    2\% \\
                           target density &                    2\% \\
               particle misidentification & \hspace*{-2.5mm}$<$1\% \\
            particle-detection efficiency & \hspace*{-2.5mm}$<$1\% \\
                 photon-beam attentuation & \hspace*{-2.5mm}$<$1\% \\
\end{tabular}
\end{ruledtabular}
\end{table}

\section{\label{section:results}Results and discussion}

In this section, we present our results for the laboratory differential cross
section obtained in this measurement for the $^3$He($\gamma,d)$ reaction at
$\theta^{\rm LAB}=90^\circ$, and we compare these results to previous data and
calculations.  The differential cross-section data are summarized in 
Table \ref{table:data}.

\begin{table}
\caption{\label{table:data}
A summary of the laboratory differential cross-section data for the 
$^3$He($\gamma,d)$ reaction measured at $\theta^{\rm LAB}=90^\circ$.  The first
uncertainty is statistical and the second uncertainty is systematic.  See also
Figs.\ \ref{figure:previous_data} and \ref{figure:calculations}.}
\begin{ruledtabular}
\begin{tabular}{cr}
$E_{\gamma}$ &   $\frac{d\sigma}{d\Omega}(\theta^{\rm LAB}=90^\circ)$ \\
       (MeV) &                   ($\mu$b/sr) \\
\hline
        14.0 &     85.0 $\pm$ 5.3 $\pm$ 7.2 \\
        14.9 &     94.4 $\pm$ 7.6 $\pm$ 9.4 \\
        15.7 &     82.3 $\pm$ 3.3 $\pm$ 7.2 \\
        16.7 &     77.0 $\pm$ 8.2 $\pm$ 7.7 \\
        17.7 &     71.4 $\pm$ 2.8 $\pm$ 6.4 \\
        19.9 &     59.0 $\pm$ 1.8 $\pm$ 3.2 \\
        24.0 &     47.8 $\pm$ 2.6 $\pm$ 2.4 \\
        25.6 &     42.0 $\pm$ 2.7 $\pm$ 2.1 \\
        26.2 &     41.3 $\pm$ 1.8 $\pm$ 1.7 \\
        28.8 &     34.3 $\pm$ 1.6 $\pm$ 1.1 \\
        31.2 &     28.1 $\pm$ 1.3 $\pm$ 0.9 \\
\end{tabular}
\end{ruledtabular}
\end{table}

\subsection{\label{subsection:calculations}The calculations}

Both the calculations of Skibi\'nski {\it{et al.}} \cite{Ski03} and Deltuva 
{\it{et al.}} \cite{Del05} to which we compare our data employed the Faddeev 
technique.  Further, they both include final-state interactions.  The authors 
chose different nuclear potentials and treated the two-body nuclear-current 
operator differently.  Further, their treatments of the Coulomb interaction
were not the same. Skibi\'nski {\it{et al.}} considered it only in the bound 
state and not in the continuum.  
This is believed to increase the predicted cross section for energies close to 
threshold, with a negligible effect above 15 MeV \cite{Gol02}.  Deltuva 
{\it{et al.}} included the Coulomb interaction in both the bound and the 
scattering state. This was accomplished using the screening and renormalization
approach described in detail in Ref.\ \cite{Del05}.

In order to isolate 3NF effects, Skibi\'nski {\it{et al}}.\ calculated the 
cross section using the NN potential AV18 both with and without the explicit 
inclusion of the Urbana IX 3NF.  The two-body nuclear-current operator was 
included using the Siegert theorem \cite{Sie37,Sac51,Fri84}.  
The authors found that the binding energy 
of $^{3}$He decreased from $-$6.92 MeV (AV18 alone) to $-$7.74 MeV with the 
inclusion of the Urbana IX 3NF, in good agreement with the experimental value
of $-$7.72 MeV\@.

Deltuva {\it{et al}}.\ used the CD-Bonn NN potential together with the 
coupled-channel CD-Bonn+$\Delta$.  The authors claim that the inclusion of 
the $\Delta$-isobar corresponds to the implicit inclusion of a 3NF.  The 
Siegert theorem, together with explicit inclusion of one and two-body currents 
not accounted for by the theorem, was used for the two-body nuclear-current 
operator.  Despite the inclusion of the non-Siegert terms, current conservation
is not fulfilled.  The predictions are essentially insensitive to the 
inclusion of 3NF effects.  These authors found that the binding energy of 
$^{3}$He decreased from $-$7.26 MeV (CD-Bonn alone) to $-$7.54 MeV with the 
inclusion of the $\Delta$, somewhat higher than the experimental value of 
$-$7.72 MeV\@.

\subsection{\label{subsection:comparisons}Comparison to previous data}

\begin{figure}
\begin{center}
\resizebox{0.5\textwidth}{!}{\includegraphics{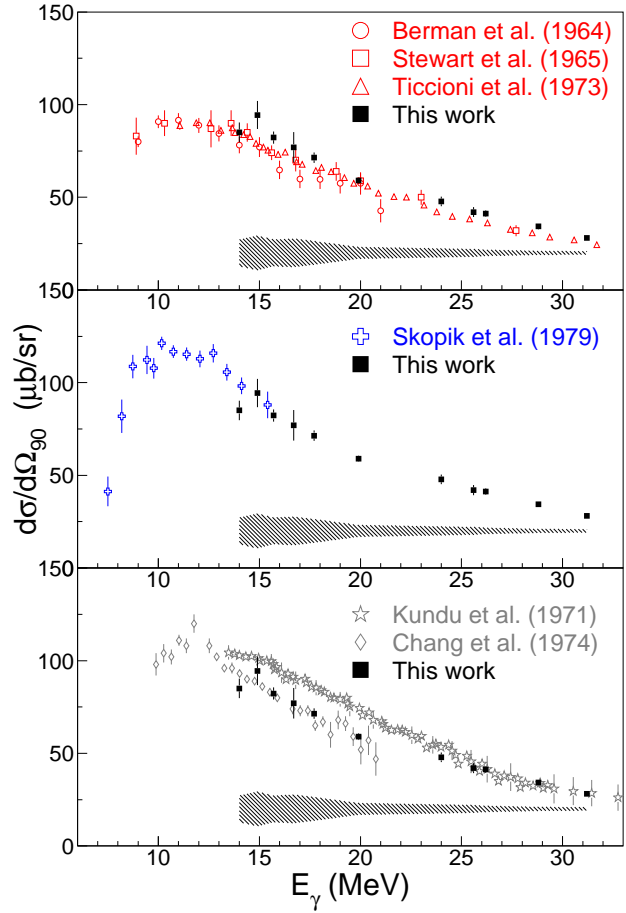}}
\caption{\label{figure:previous_data}(Color online) The laboratory differential
cross section obtained in this measurement for the $^3$He($\gamma,d)$ reaction 
at $\theta^{\rm LAB}=90^\circ$ (solid black squares) compared to previous 
results.  In every case, error bars are the statistical uncertainties, while 
the systematic uncertainty in this measurement is represented by the bands at 
the base of the panels.  The top panel presents a comparison of our 
tagged-photon data to previous $(\gamma,d)$ measurements; the middle panel 
presents a comparison to a previous $(p,\gamma)$ measurement; and the bottom 
panel presents a comparison to previous $(e,d)$ measurements which have been 
expressed at the real-photon point.  See text for details.}
\end{center}
\end{figure}

Figure \ref{figure:previous_data} shows the laboratory differential 
cross-section data obtained in this measurement for the $^3$He($\gamma,d)$ 
reaction at $\theta^{\rm LAB}=90^\circ$ (solid black squares) compared to 
previous results.  In every case, error bars are the statistical uncertainties,
while the systematic uncertainty in this measurement is represented by the 
bands at the bases of the panels.  The top panel presents a comparison of our 
tagged-photon data to previous bremsstrahlung $(\gamma,d)$ 
measurements \cite{Ber64,Ste65,Tic73}; the middle panel presents a comparison 
to a previous $(p,\gamma)$ measurement \cite{Sko79}; and the bottom panel 
presents a comparison to previous $(e,d)$ measurements \cite{Kun71,Cha74} which
have been converted to the real-photon point using ``virtual-photon theory" -- 
see the aforementioned Refs.\ for details. 
For readability, only selected error bars have been plotted on the Kundu 
{\it et al.} data.  Further, note that while the Stewart {\it et al.}, Ticcioni 
{\it et al.}, and Kundu {\it et al.} data sets all extend above 35 MeV, these 
higher-energy data are not shown here.

It is clear that the previous data all agree reasonably well with the present 
results. The bremsstrahlung $(\gamma,d)$ measurements shown in the top panel 
all fall systematically slightly below our tagged-photon results.  We note 
that Berman {\it et al.}, Stewart {\it et al.}, and Ticcioni {\it et al.} 
claim systematic 
uncertainties of 6, 10, and 6\%, respectively.  The $(p,\gamma)$ measurement 
seems to ``mesh" with our data, especially when systematic uncertainties are 
considered.  Skopik {\it et al.} claim a systematic uncertainty of 10\%.
However, in the vicinity of 15 MeV, no general trend is readily apparent, and 
our data do not extend low enough in energy to draw general 
conclusions \footnote{
The present measurement actually consisted of two complementary detector 
setups which extended over a large photon-energy range.  The results obtained 
using the setup detailed in this paper constituted the higher tagged-photon
energy region.  These data were acquired to overlap with the previously 
mentioned SAL \cite{Ori09} measurement.  A second detector setup consisting of 
a Bragg/PPAC chamber \cite{Kot99} was employed to investigate lower 
tagged-photon energies extending down to at least 12 MeV (worst-case scenario).
These data were being analyzed by our colleague A.~A. Kotov at the time of his 
death, and are not ready for publication.}.
The scatter in the electrodisintegration data expressed at the real-photon 
point shown in the bottom panel is the largest, and we note that the 
photon-energy dependence of these two $(e,d)$ measurements is very different.  
This may be due to normalization issues with one or both measurements, or the 
inaccuracy of virtual-photon theory for photon energies significantly lower 
than the endpoint of the bremsstrahlung spectrum, as suggested by Chang 
{\it et al.} in \mbox{Ref.\ \cite{Cha74}}.  In the energy region below 15 MeV, 
the Kundu {\it et al.} cross section rises slowly with decreasing photon 
energy, perhaps reaching a maximum at around
13 MeV and ``turning over" at lower photon energies. This is in contrast to the
Chang {\it et al.} results which rise sharply.  Thus, the aforementioned 
disagreement between the electrodisintegration data sets is not just due to 
systematic scaling.
The Kundu {\it et al.} data agree with ours for $E_{\gamma}$ $>$ 25 MeV, while 
the Chang {\it et al.} data agree with ours for $E_{\gamma}$ $<$ 20 MeV\@.  We 
note that Kundu {\it et al.} and Chang {\it et al.} claim systematic 
uncertainties of 15\% and 6\%, respectively.

\begin{figure}
\begin{center}
\resizebox{0.5\textwidth}{!}{\includegraphics{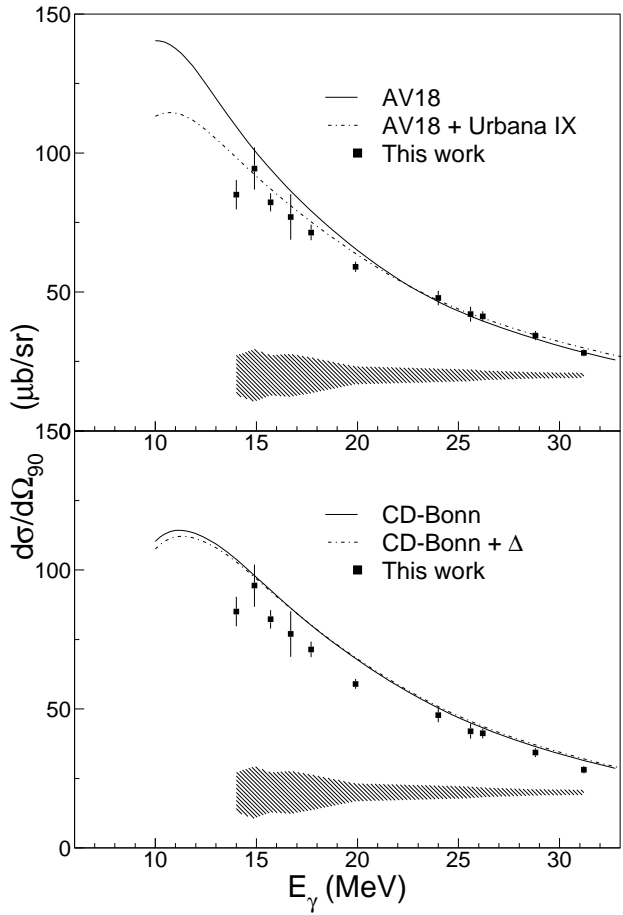}}
\caption{\label{figure:calculations}The laboratory differential cross section 
obtained in this measurement for the $^3$He($\gamma,d)$ reaction at 
$\theta^{\rm LAB}=90^\circ$ (solid black squares) compared to theoretical
predictions.  Error bars are the statistical uncertainties, while the 
systematic uncertainty is represented by the bands at the base of the panels.
The top panel presents a comparison to the predictions of Skibi\'nski 
{\it{et al}}, while the bottom panel presents a comparison to the predictions
of Deltuva {\it{et al}}.  See text for details.}
\end{center}
\end{figure}

Figure \ref{figure:calculations} shows the laboratory differential cross 
section obtained in this measurement for the $^3$He($\gamma,d)$ reaction at
$\theta^{\rm LAB}=90^\circ$ (solid black squares) compared to the previously 
discussed theoretical predictions.  Again, error bars are the statistical 
uncertainties, while the systematic uncertainty is represented by the bands at 
the bases of the panels.  The top panel presents a comparison to the 
predictions of Skibi\'nski {\it{et al.}}, while the bottom panel presents a 
comparison to the predictions of \mbox{Deltuva {\it{et al}}}.

Both calculations do a reasonable job of predicting the present results over
the energy range of the experiment.  That said, the improved accuracy and
precision of the present data may be exploited to distinguish between the
four cases presented in Fig.\ \ref{figure:calculations}.  As shown in the top 
panel, within the calculational framework of Skibi\'nski {\it{et al.}}, our 
data favor the inclusion of 3NF.  As shown in the bottom panel, within the 
calculational framework of Deltuva {\it{et al.}}, no conclusions regarding the 
inclusion of 3NF may be drawn, as the calculational sensitivity is very small.
We thus see that at all energies, including those above 23 MeV where our
systematic uncertainties are the smallest (roughly 5\%), the present data 
favor the calculation of Skibi\'nski {\it{et al.}} that includes 3NF.

\section{Summary and conclusions}

In summary, the differential cross section for the two-body photodisintegration
of $^3$He has been measured at $\theta^{\rm LAB}=90^\circ$ using tagged photons
in the energy range 14 -- 31 MeV, and the cross-section data have been compared
to the results of other available measurements and theoretical calculations.  

Most of the previous data sets agree reasonably well with the present results. 
The previous bremsstrahlung $(\gamma,d)$ measurements \cite{Ber64,Ste65,Tic73} 
all report differential cross sections which are systematically smaller than 
our cross section.  The previous $(p,\gamma)$ measurement \cite{Sko79} 
agrees reasonably well with our results in the limited region of overlap.  
However, this region of overlap is so small that it is difficult to draw 
general conclusions.  The reported electrodisintegration cross sections   
expressed at the real-photon point \cite{Kun71,Cha74} are generally larger 
than our cross section.  That said, there appears to be a disagreement between 
these two data sets 
which is not simply due to systematic scaling, as they represent completely 
different excitation functions.  This disagreement may be rooted in the
inaccuracy of virtual-photon theory for photon energies significantly lower
than the endpoint of the bremsstrahlung spectrum.

Theoretical predictions based upon Faddeev calculations using realistic 
potentials and which take into account 3NF and FSI compare favorably with our 
tagged-photon data, especially considering systematic uncertainties. Our data, 
the previous data for the $^3$He$(\gamma,d)$ reaction \cite{Ber64,Ste65,Tic73},
and the previous data for the $(p,\gamma)$ reaction \cite{Sko79} favor the 
calculational framework of Skibi\'nski {\it{et al.}} \cite{Ski03} with the 
inclusion of 3NF.  Clearly, photodisintegration studies of the three-nucleon 
system should be continued at photon energies below 20 MeV as well as above 
70 MeV where the calculations of Skibi\'nski {\it{et al.}} \cite{Ski03b} show 
an enhanced sensistivity to 3NF effects.

We direct the interested reader to Refs. \cite{Ann04,Fis08,Fis09} 
for an overview of a newly commenced program of experiments at the
recently upgraded Tagged-Photon Facility at MAX-lab.  This program consists
of a systematic investigation of the photodisintegration of $^{3,4}$He 
using tagged photons and a novel gas-scintillator active target developed at 
the University of Glasgow, UK to detect heavy charged recoil fragments down 
to the reaction threshold in combination with standard external detectors to 
detect ejected neutrons, protons, and deuterons, as well as scattered photons.

\begin{acknowledgments}

The authors acknowledge the outstanding support of the MAX-lab staff which made
this experiment successful.  The Lund group acknowledges the financial support 
of the Swedish Research Council, the Knut and Alice Wallenberg Foundation, the 
Crafoord Foundation, the Swedish Institute, the Wenner-Gren Foundation, and the
Royal Swedish Academy of Sciences.  This work was sponsored in part by the 
U.S. Department of Energy under grants DE-FG02-95ER40901 and DE-FG02-99ER41110.
Partial support was also provided by Jefferson Lab via the Southeastern 
Universities Research Association under U.S. Department of Energy grant 
DE-AC05-84ER40150.  We thank \mbox{R. Skibi\'nski} and A. Deltuva for sharing 
their calculations with us and for general guidance.  We thank H. Griesshammer
for constructive suggestions.  We dedicate this work to the memory of our 
colleague A. A. Kotov.

\end{acknowledgments}

\bibliography{karlsson_etal}

\end{document}